\documentclass[10pt]{article}
\usepackage{graphicx}
\def\Title#1{\begin{center} {\Large #1 } \end{center}}
\def\Author#1{\begin{center}{ \sc #1} \end{center}}
\def\Address#1{\begin{center}{ \it #1} \end{center}}

\newcommand\pubblock{\rightline{\begin{tabular}{l} Proceedings of the Fifth Annual LHCP\\ \pubnumber\\
         \pubdate  \end{tabular}}}

\newenvironment{Abstract}{\begin{quotation} \begin{center} 
             \large ABSTRACT \end{center}\bigskip 
      \begin{center}\begin{large}}{\end{large}\end{center} \end{quotation}}

\newenvironment{Presented}{\begin{quotation} \begin{center} 
             PRESENTED AT\end{center}\bigskip 
      \begin{center}\begin{large}}{\end{large}\end{center} \end{quotation}}

\def\beq{\begin{equation}}
\def\eeq#1{\label{#1}\end{equation}}
\def\eeqn{\end{equation}}

\def\beqa{\begin{eqnarray}}
\def\eeqa#1{\label{#1}\end{eqnarray}}
\def\eeqan{\end{eqnarray}}

\let\bar=\overbar

\def\Dslash{\not{\hbox{\kern-4pt $D$}}}
\def\dslash{\not{\hbox{\kern-2pt $\del$}}}

\def\msb{{\bar{\ssstyle M \kern -1pt S}}}

\usepackage{color}
\usepackage{url}

\newcommand\pubnumber{ ATL-PHYS-PROC-2017-XXX }

\newcommand\pubdate{\today}

\newcommand{\ifb}{\ensuremath{\mathrm{fb^{-1}}}}

\newcommand{\zlljets}{\ensuremath{Z(\rightarrow \ell \ell)}}
\newcommand{\SHERPA} {{\textsc{sherpa}}}
\newcommand{\MADGRAPH} {\textsc{MadGraph}}
\newcommand{\PYTHIA} {{\textsc{pythia}}}
\newcommand{\Zc}   {\ensuremath{Z + c}}
\newcommand{\Zb}   {\ensuremath{Z + b}}
\newcommand{\POWHEG} {{\textsc{powheg}}}
\newcommand{\MCATNLO} {\textsc{mc@nlo}}
\newcommand{\MGvATNLO}{\MADGRAPH{}5\_a\MCATNLO}
\newcommand{\MGvATMC}{\textsc{MG}5\_aMC}
\newcommand{\BLACKHAT} {\textsc{BlackHat}}
\def\affiliation{
On behalf of the CMS Collaboration, \\
Institute of High Energy Physics \\
Chinese Academy of Sciences, Beijing, 100049, China}

\begin{document}

\large
\begin{titlepage}
\pubblock

\vfill
\Title{Measurements of associated production of vector bosons and jets in CMS}
\vfill

\Author{Muhammad Ahmad}
\Address{\affiliation}
\vfill
\begin{Abstract}
The most recent results of Standard Model physics using 8 and 13 TeV proton-proton collision data recorded by the CMS detector during the 
LHC Runs I and II are reviewed. This overview includes studies of several results of vector boson production in association with jets. The 
outlined results are compared to the corresponding theoretical predictions and no significant deviation is observed.

\end{Abstract}
\vfill

\begin{Presented}
The Fifth Annual Conference\\
 on Large Hadron Collider Physics \\
Shanghai Jiao Tong University, Shanghai, China\\ 
May 15-20, 2017
\end{Presented}
\vfill
\end{titlepage}
\def\thefootnote{\fnsymbol{footnote}}
\setcounter{footnote}{0}
%

\normalsize 


\section{Introduction}
Standard Model (SM) V+Jets results using the LHC Run I and II data have provided us a possibility to perform extensive tests of the 
electroweak and strong interactions. These measurements are made with the datasets collected by CMS 
corresponding to integrated luminosities upto 19.8 and 35.9~$\ifb$~at~$\sqrt{s} = 8$~and 13 TeV respectively.

\section{Measurement of Z+jets cross section}

\subsection{Z+jets differential cross section at 8 and 13 TeV}
Differential cross section measurements for Z boson production in association with jets at $\sqrt{s} = 8(13)$~TeV, in
the electron and muon (muon) decay channels, using a data sample corresponding to an integrated luminosity of 19.6(2.5)~\ifb~\cite{zpjet:8tev,zpjet:13tev}. The 
measurements are compared with calculations obtained from different multileg ME event generators with leading order (LO) MEs
(tree level), NLO MEs and a combination of NLO and LO MEs. 
Fig.~\ref{zpjetfig1}~shows measured differential cross sections as a function of $1^{st}$~jet transverse momentum and the
$d^2\sigma/d p_{T}({\rm j}{}_{1})d|y({\rm j}_{1})|$, here the jet absolute rapidity range up to $4.7$. While Fig.~\ref{zpjetfig2} shows the measured differential cross section as a function of transverse momentum of leading jet, the jets inclusive multiplicity and $H_T$ for inclusive jet multiplicity of one jet. The cross section measurements are also compared with the theoretical predictions. The ratios of the theoretical predictions obtained
from \MADGRAPH~5 + \PYTHIA~6, \SHERPA~2, and \MGvATMC + \PYTHIA~8 to the measurements are also shown.
\begin{figure}[htb]
\centering
\begin{minipage}[b]{0.45\textwidth}
\centering
\includegraphics[height=5.7cm]{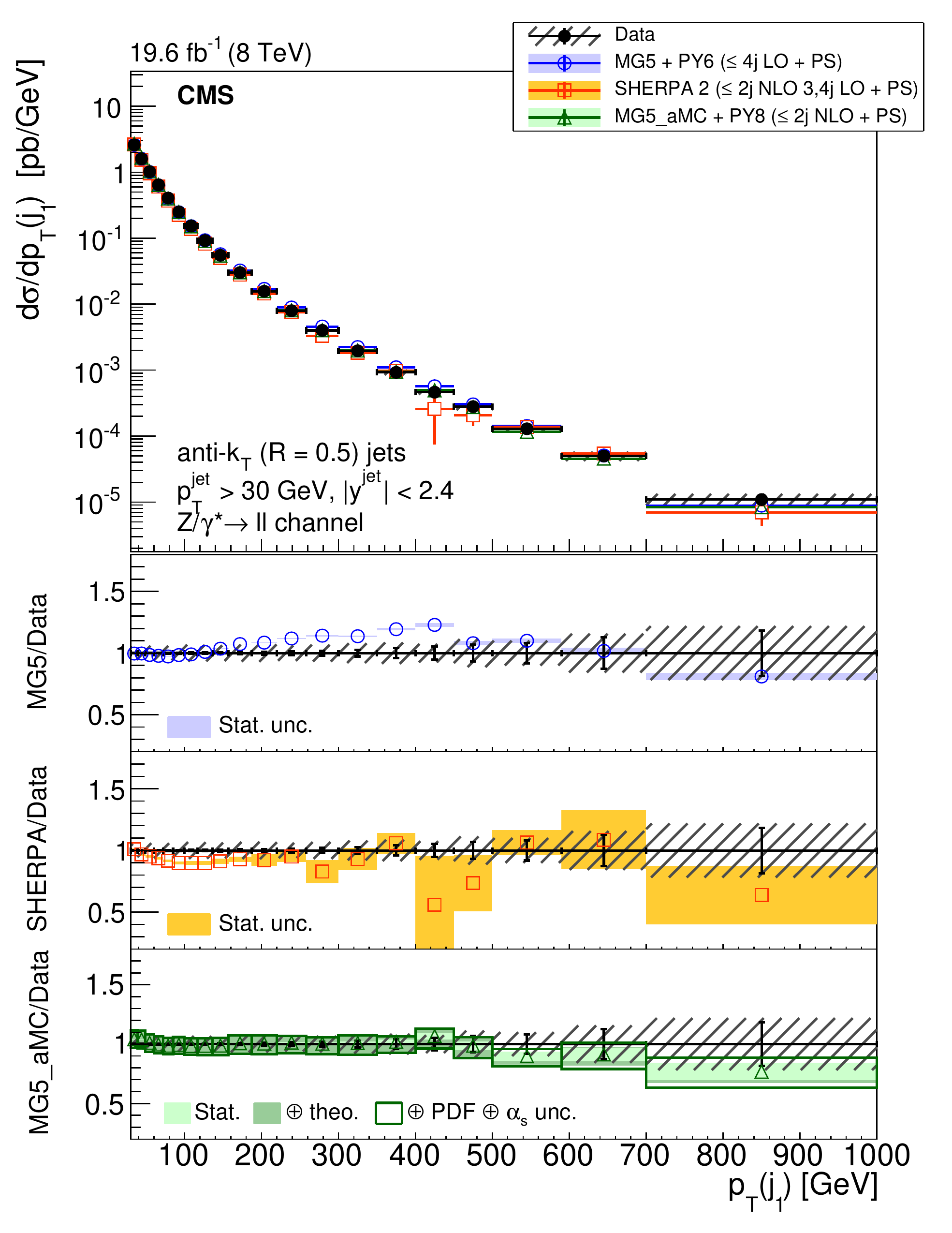}
\end{minipage}
\begin{minipage}[b]{0.45\textwidth}
\centering
\includegraphics[height=5.7cm]{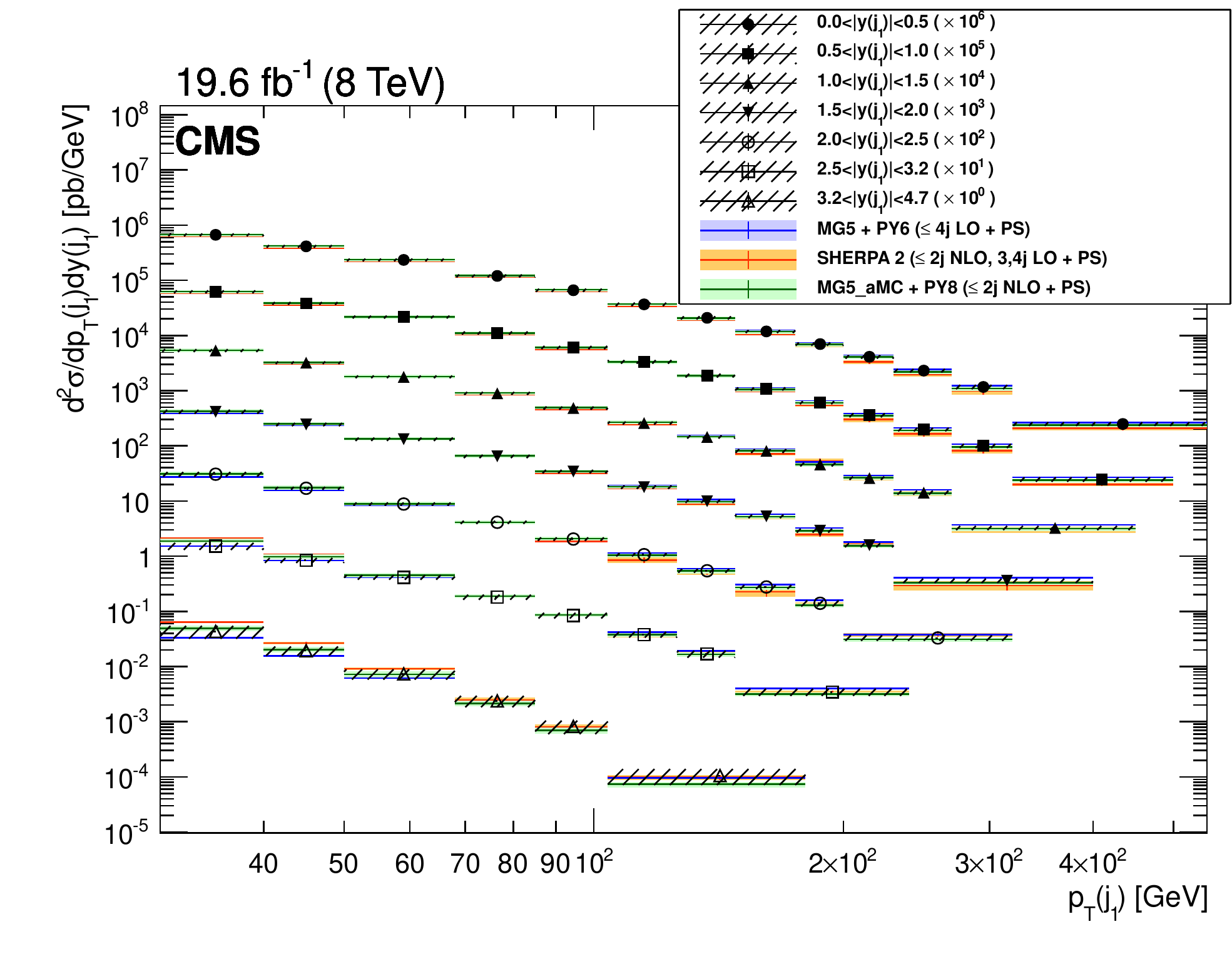}
\end{minipage}
\caption{The differential cross section for \zlljets+jets~production measured as a function of the leading jet $p_T$ (left) and leading jet transverse momentum and rapidity 
(right) compared to the predictions~\cite{zpjet:8tev}. }
\label{zpjetfig1}
\end{figure}
\begin{figure}[htb]
\centering
\begin{minipage}[b]{0.32\textwidth}
\centering
\includegraphics[height=5.7cm]{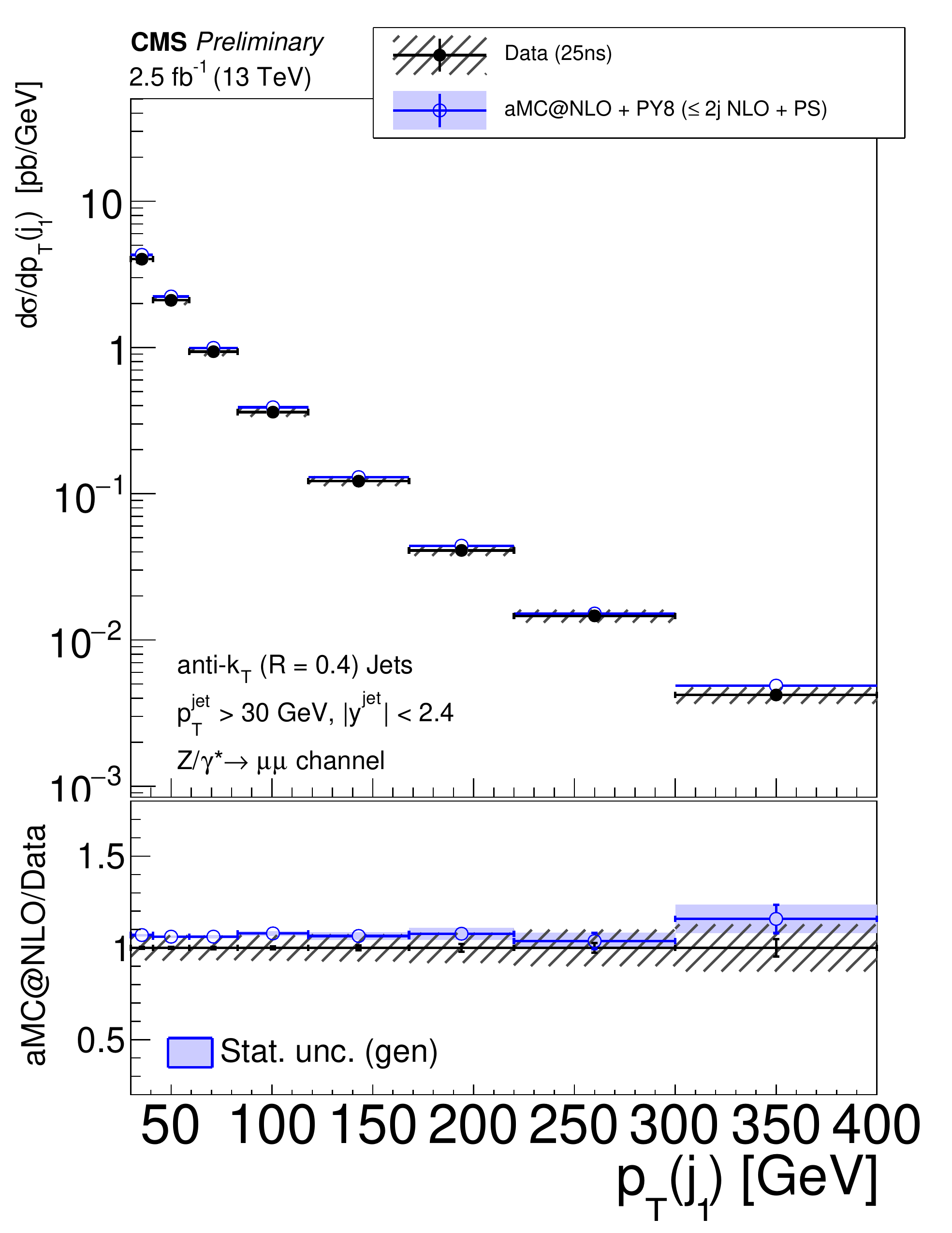}
\end{minipage}
\begin{minipage}[b]{0.32\textwidth}
\centering
\includegraphics[height=5.7cm]{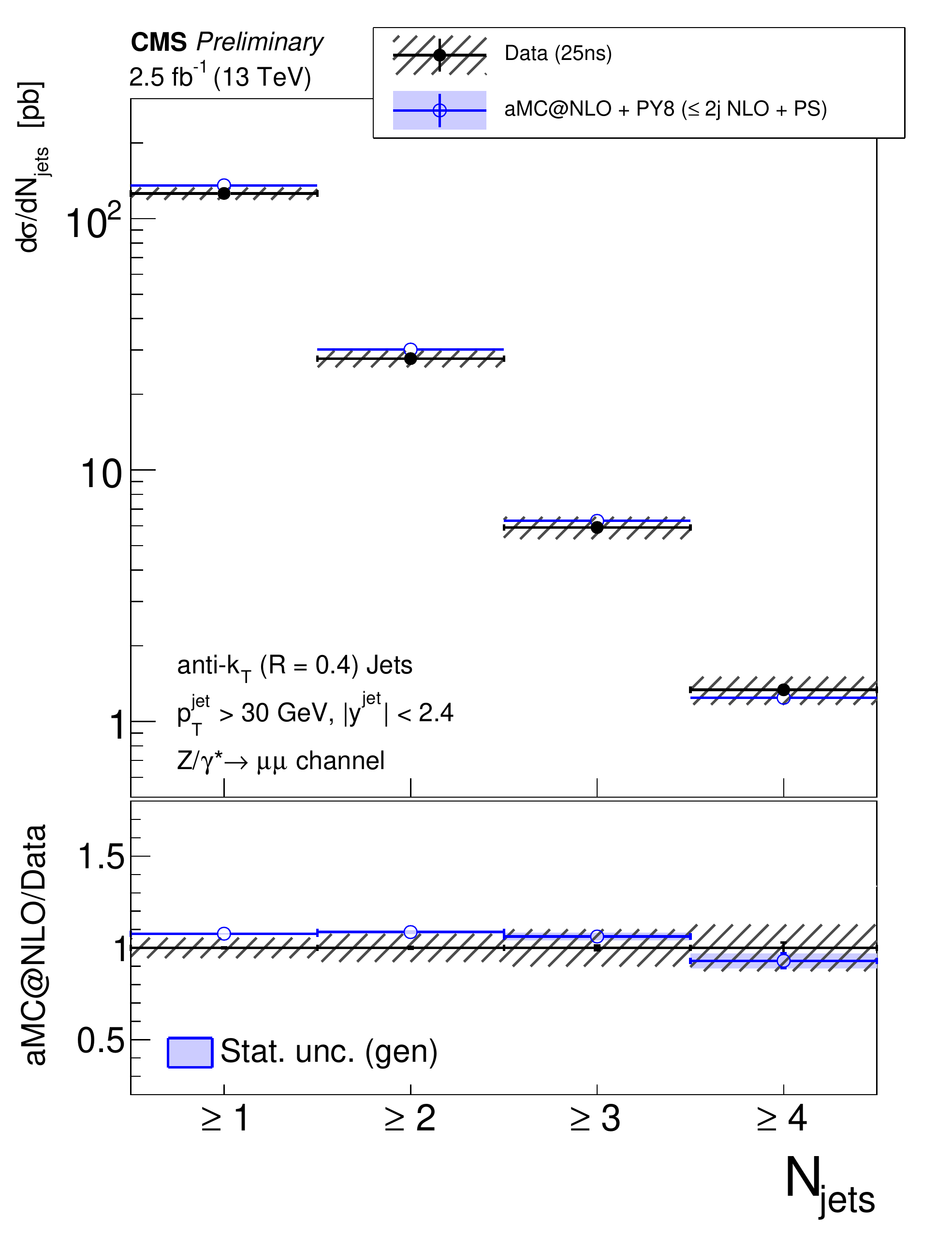}
\end{minipage}
  \begin{minipage}[b]{0.32\textwidth}
  \includegraphics[height=5.7cm]{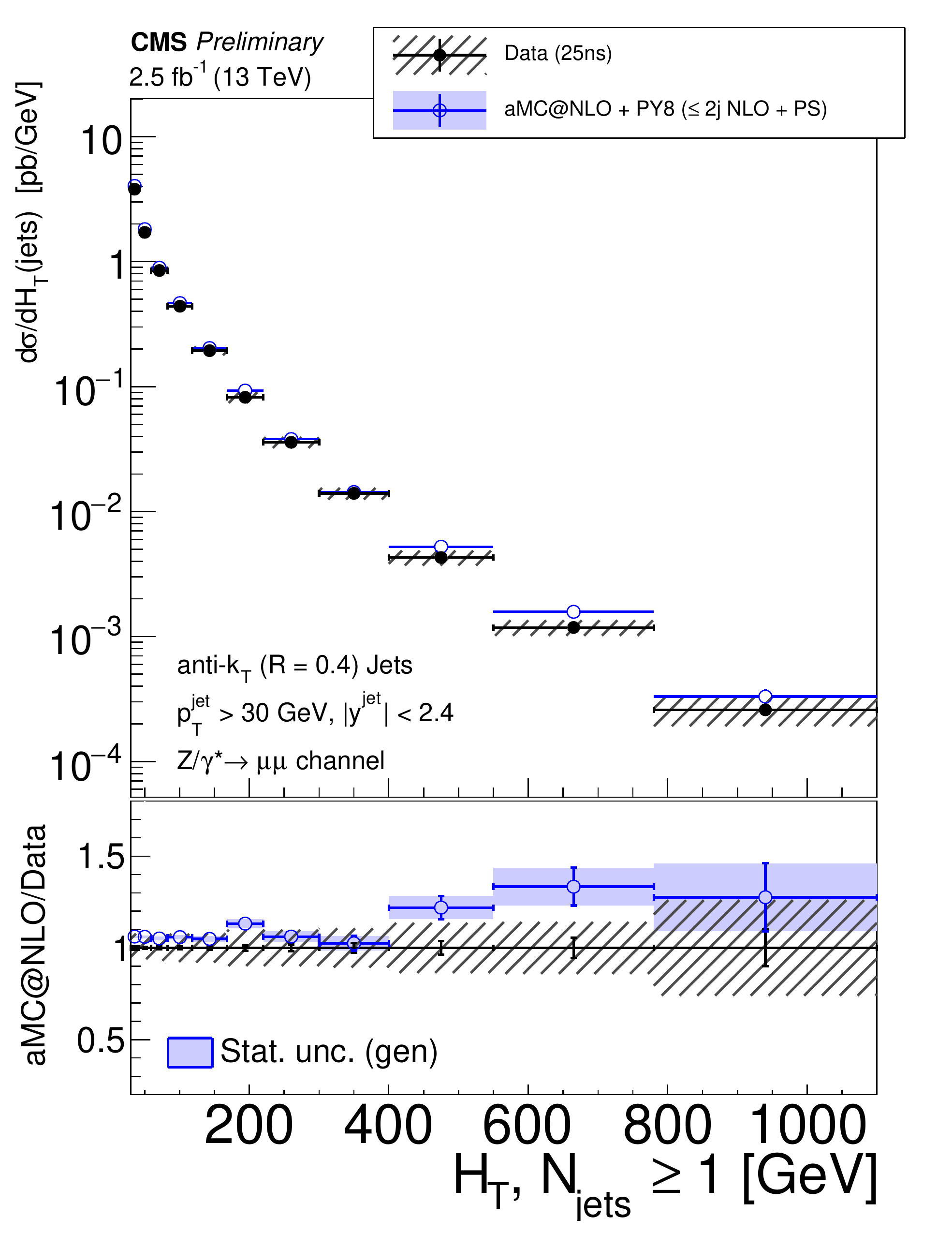}
\end{minipage}
\caption{The differential cross section for \zlljets+jets~production measured as a function of the~$p_T$~of leading jet (left), the jets inclusive multiplicity (middle) and~$H_T$~
for inclusive jet multiplicities (right) compared to the predictions~\cite{zpjet:13tev}.}
\label{zpjetfig2}
\end{figure}
\subsection{Z+b jets differential cross section at 8 TeV}
CMS Collaboration has also measured the associated production of a Z boson with at least one jet originating from a b quark in proton-proton 
collisions using~$\sqrt{s} = 8$~TeV dataset corresponding to an integrated luminosity of 19.8~\ifb~\cite{zbjet:8tev}. In this measurement, Z bosons are reconstructed through their
decays electrons and muons. The differential cross sections has been reported as a function of several observables exploiting the kinematics 
of the b jet and the Z boson. The ratios of the differential cross section for the associated production with at-least one b jet to the associated 
production with any jet are also reported. The differential cross section for the dijet system and production of Z boson with two b jets is also measured. Results are 
compared with theoretical prediction based on two different flavour schemes for the choice of initial-state partons. Selected differential fiducial cross section results for  
Z(1b) production as function of leading b jet and~$\Delta\phi_{Z b}$~are shown in Fig.~\ref{zpjetfig3}.
\begin{figure}[htb]
\centering
\begin{minipage}[b]{0.45\textwidth}
\centering
\includegraphics[height=5.7cm]{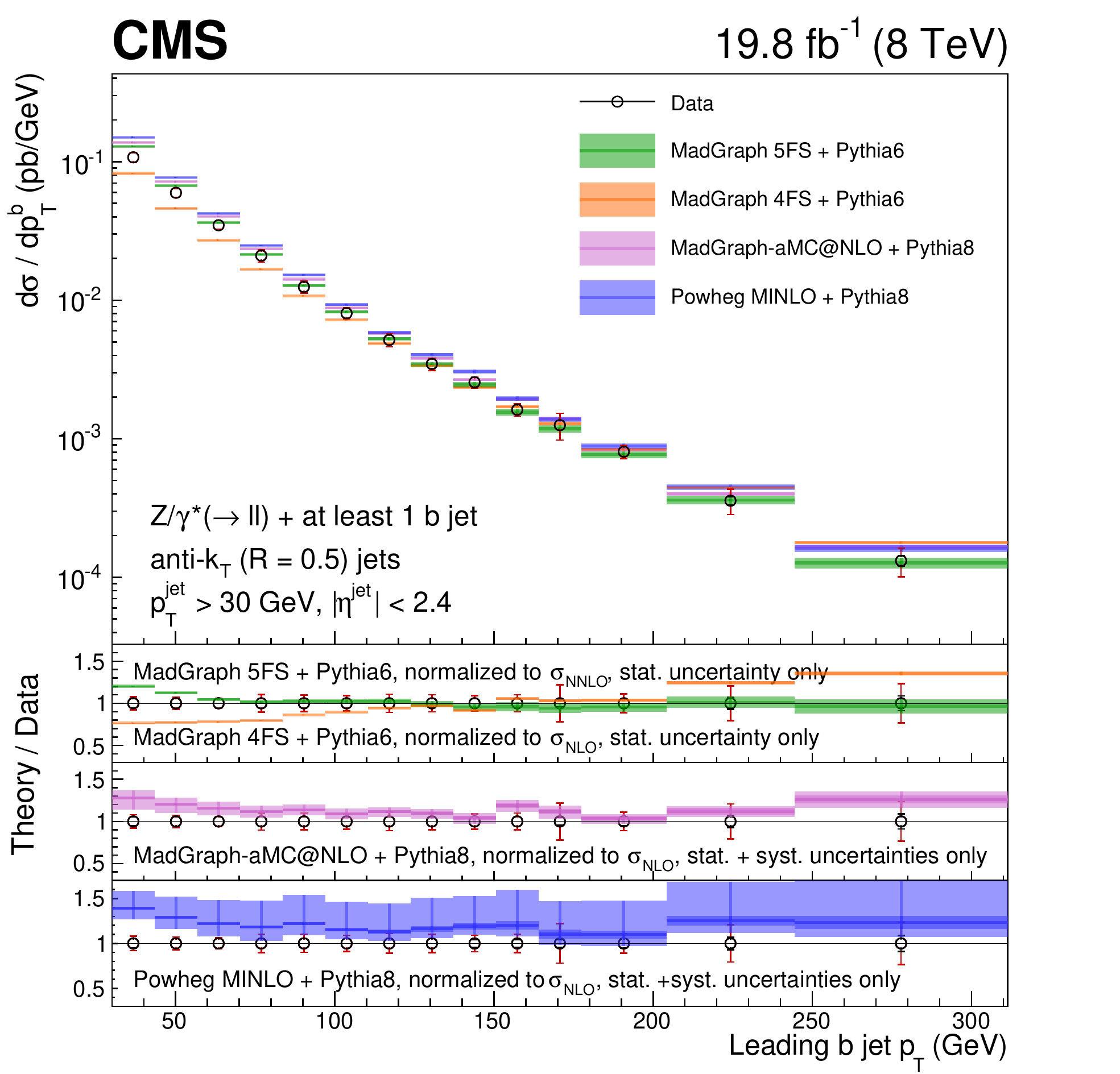}
\end{minipage}
\begin{minipage}[b]{0.45\textwidth}
\centering
\includegraphics[height=5.7cm]{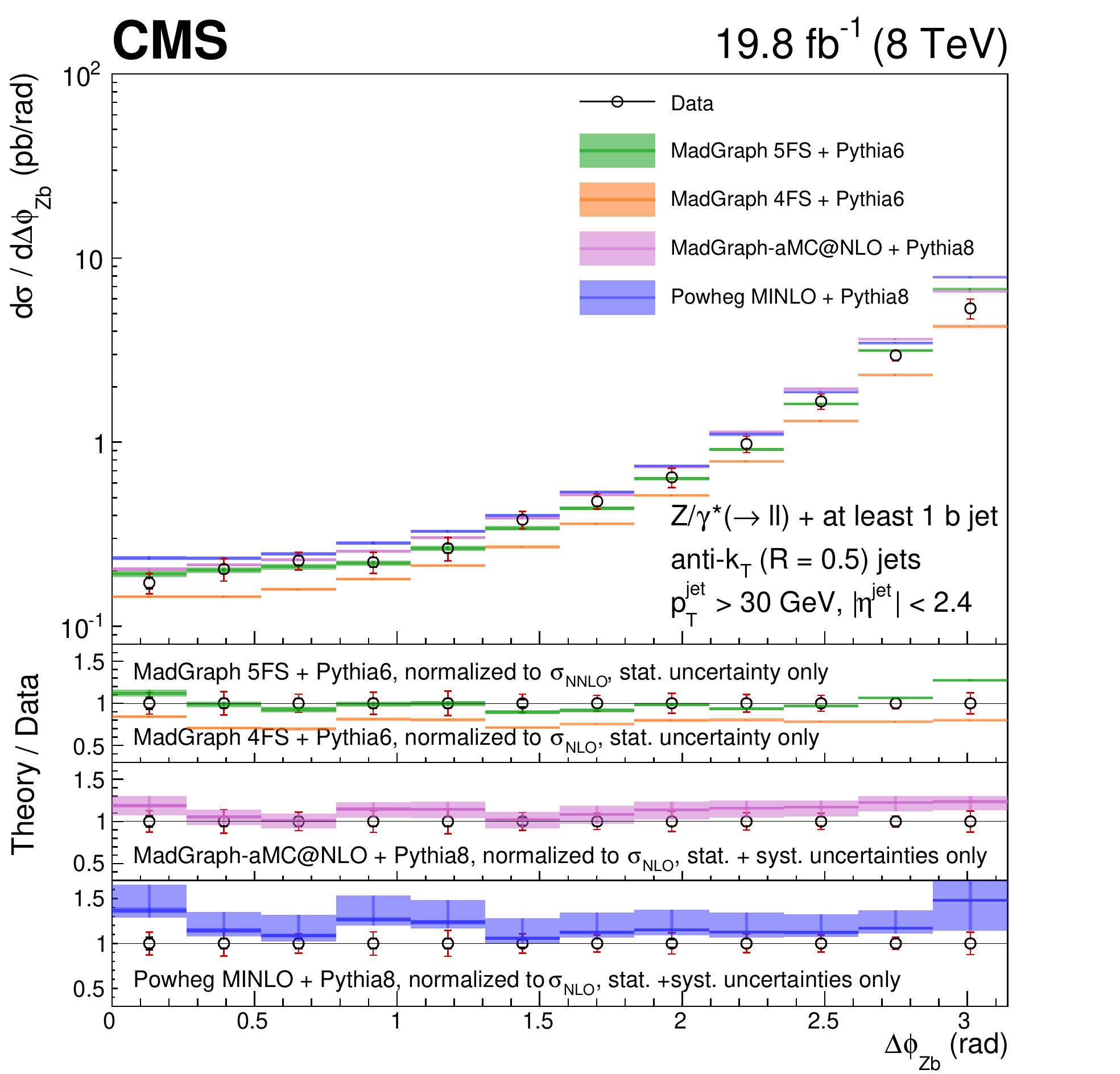}
\end{minipage}
\caption{ Differential fiducial cross section for Z(1b) production as a function of the leading b jet (left) and $\Delta\phi_{Z b}$ (right) compared with 
the \MADGRAPH 5FS, \MADGRAPH 4FS, \MGvATNLO, and \POWHEG\ \textsc{minlo} theoretical predictions (shaded bands), normalized to the theoretical cross sections~\cite{zbjet:8tev}.}
\label{zpjetfig3}
\end{figure}
\subsection{Z+c jets differential cross section at 8 TeV}
The production of a Z($\rightarrow \ell\ell$) boson (where~$\ell$~= e or~$\mu$) and a charm-quark jet (Z + c) in proton-proton collisions at $\sqrt{s} = 8$~TeV is reported in the Reference~\cite{zcjet:8teV}. The differential production cross section of~\Zc~and ratio of~$\Zc$~and~$\Zb$~are measured with data collected by the CMS experiment corresponding to an integrated luminosity of 19.7~\ifb. Jets originating from heavy flavour quarks are identified using semileptonic decays of c- or b-flavoured hadrons and hadronic decays of charm hadrons. Differential cross section results are reported as a function of transverse momentum of the Z boson and of the heavy flavour jet in Fig.~\ref{zpjetfig4}. 
\begin{figure}[htb]
\centering
\begin{minipage}[b]{0.45\textwidth}
\centering
\includegraphics[height=5.7cm]{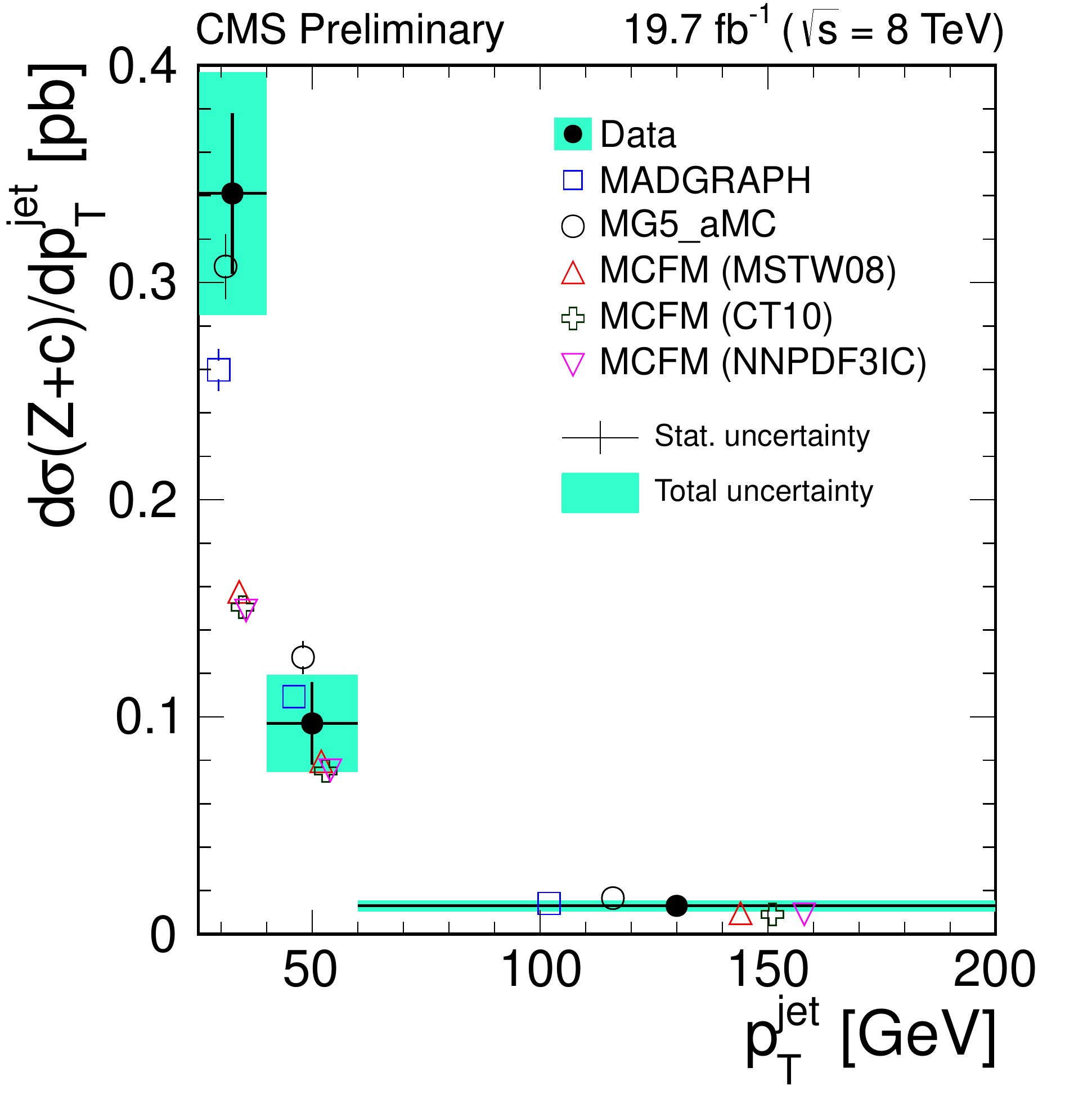}
\end{minipage}
\begin{minipage}[b]{0.45\textwidth}
\centering
\includegraphics[height=5.7cm]{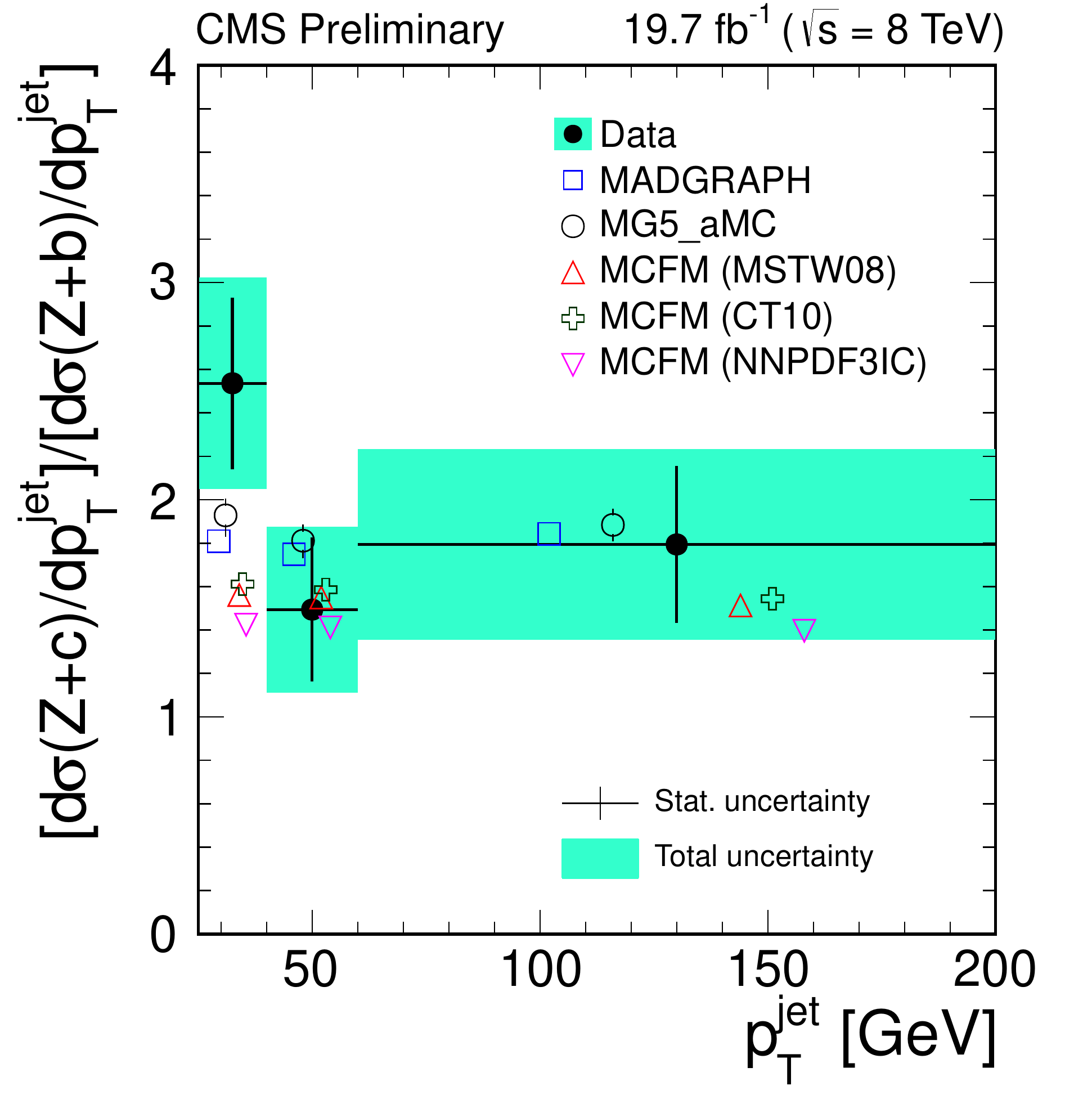}
\end{minipage}
\caption{ Differential $\Zc$ cross section (left) and $(\Zc)/(\Zb)$ cross sections ratio (right)
as a function of the transverse momentum of the jet. Statistical uncertainties in the data are shown as error bars. The solid rectangles indicate the total experimental uncertainty~\cite{zbjet:8tev}.}
\label{zpjetfig4}
\end{figure}

\section{Measurements of W+jets cross section}
\subsection{W+jets differential cross section at 8 and 13 TeV}
Differential cross sections for a~$\mathrm{W}(\rightarrow \mu\nu)$~boson in association with jets is presented in the References~\cite{wpjet:8tev,wpjet:13tev}. The measurement
 is based on the~$13(8)~\mathrm{TeV}$~proton-proton collisions data corresponding to integrated luminosity of~$2.5 (19.6)~\mathrm{fb}^{-1}$ recorded by the CMS detector. The 
cross sections are reported as a function of jet multiplicity, the jet transverse momenta and the scalar sum of the jet transverse momenta for different jet multiplicities as 
shown in Fig.~\ref{zwpjetfig1}~and~\ref{zwpjetfig2}. The measured cross sections are compared with the predictions that include multileg leading order and next-to-leading order matrix element calculations interfaced with parton showers and a next-to-next-to-leading order calculation for W + 1 jet.
\begin{figure}[htb]
\centering
\begin{minipage}[b]{0.45\textwidth}
\centering
\includegraphics[height=5.7cm]{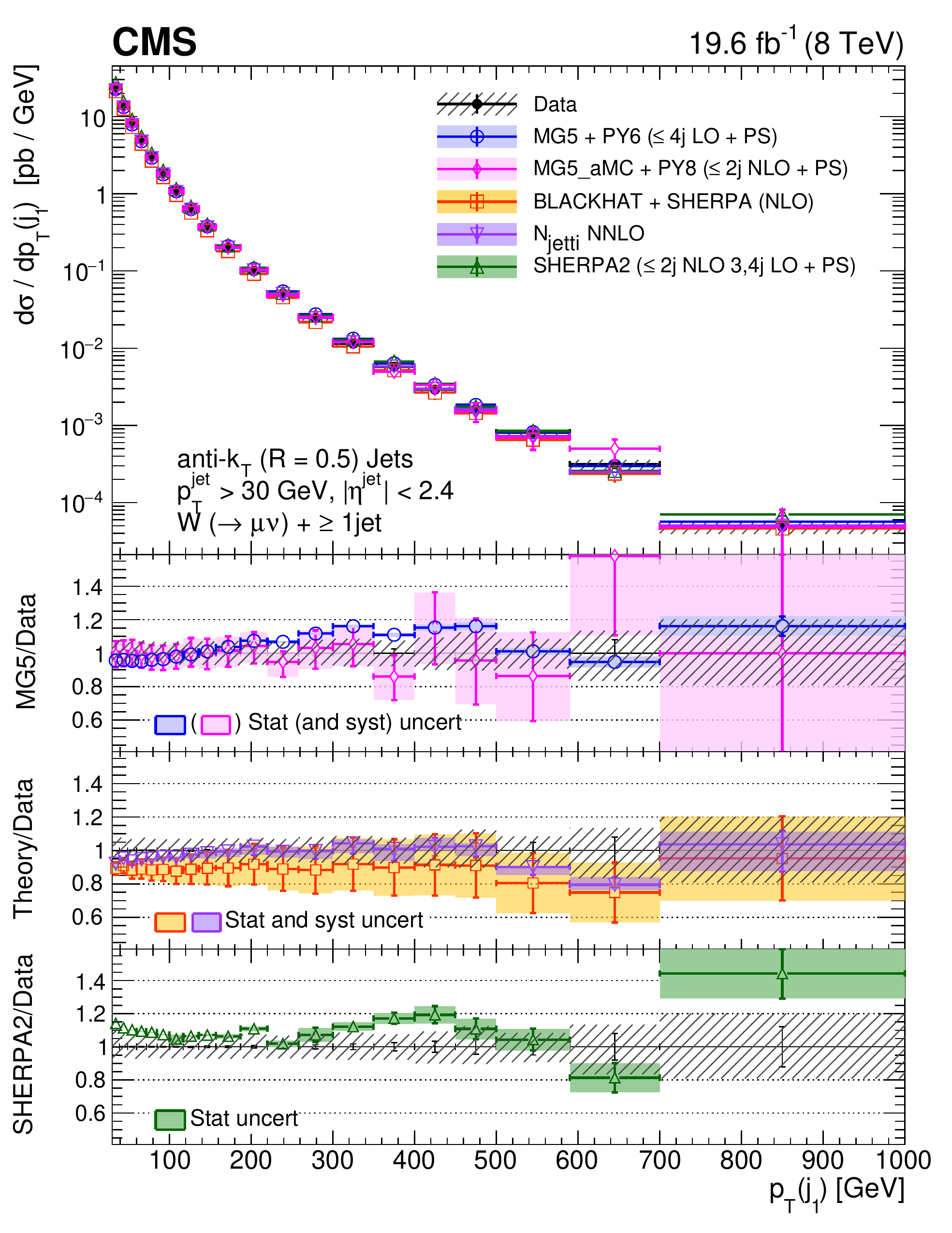}
\end{minipage}
\begin{minipage}[b]{0.45\textwidth}
\centering
\includegraphics[height=5.7cm]{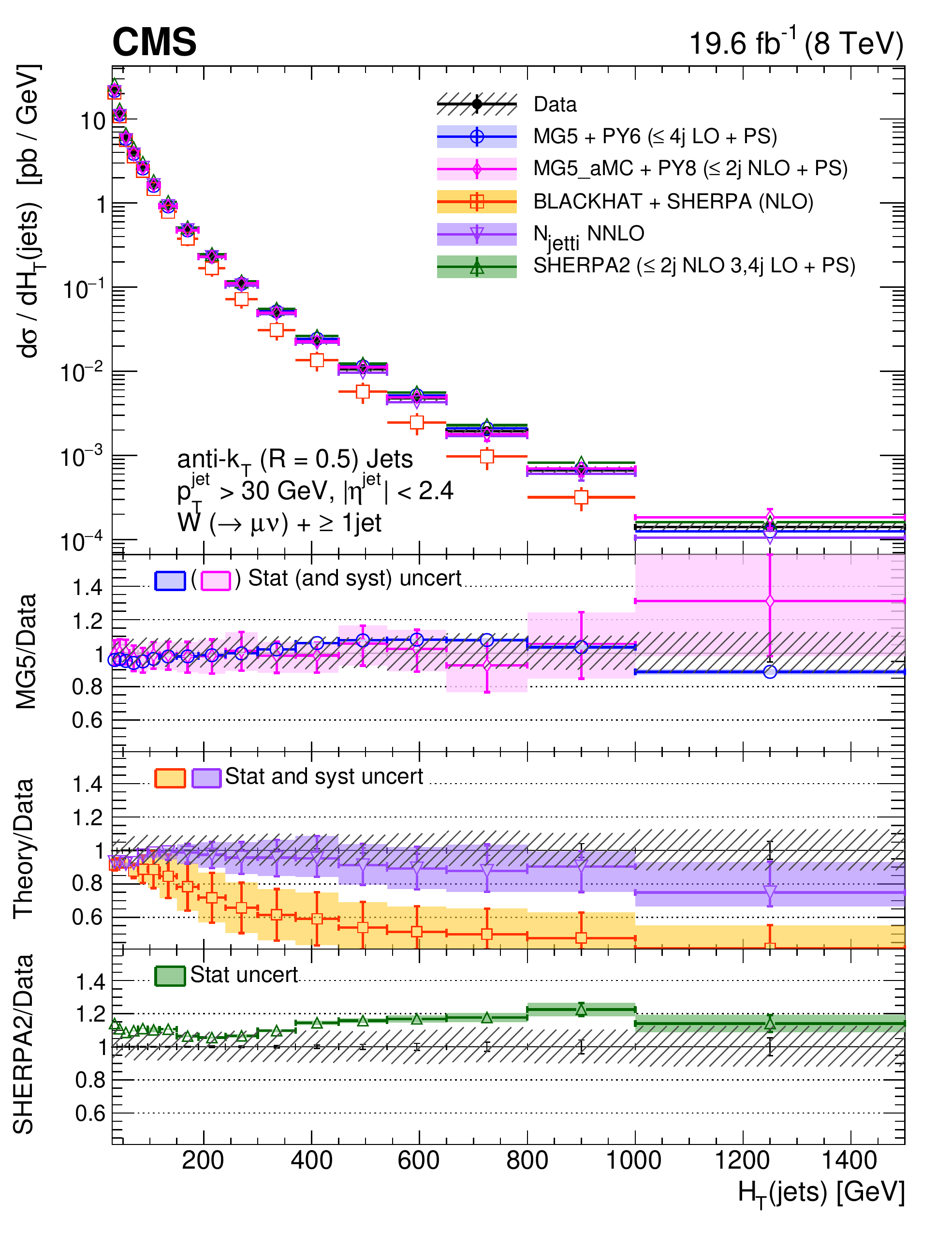}
\end{minipage}
\caption{Differential cross sections as a function of leading jet (left) and $H_T$ for inclusive jet multiplicities 1(right), compared to the predictions of \MADGRAPH, \MGvATNLO, {\SHERPA 2}, \BLACKHAT{}+\SHERPA~\cite{wpjet:8tev}.}
\label{zwpjetfig1}
\end{figure}
\begin{figure}[htb]
\centering
\begin{minipage}[b]{0.32\textwidth}
\centering
\includegraphics[height=5.7cm]{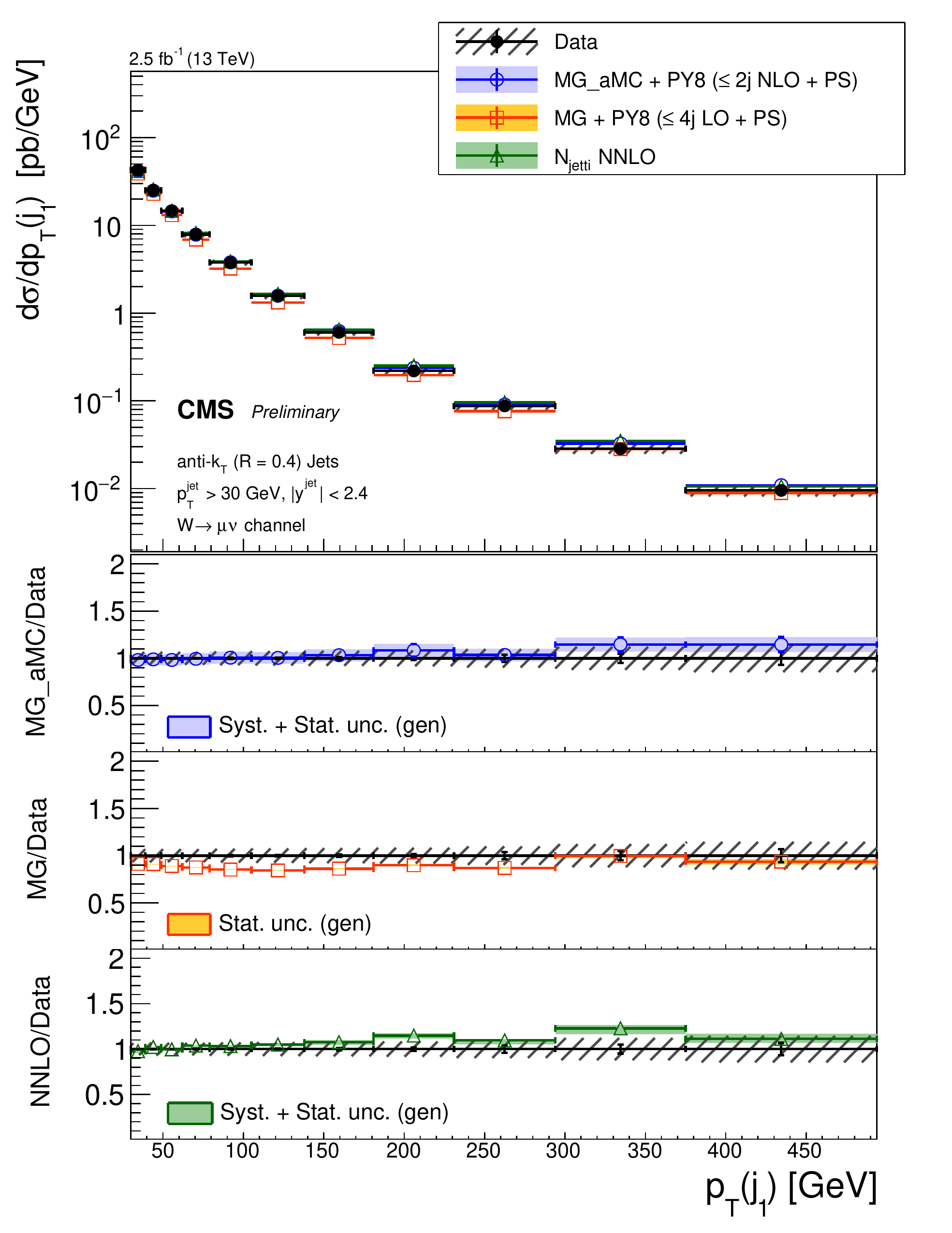}
\end{minipage}
\begin{minipage}[b]{0.32\textwidth}
\centering
\includegraphics[height=5.7cm]{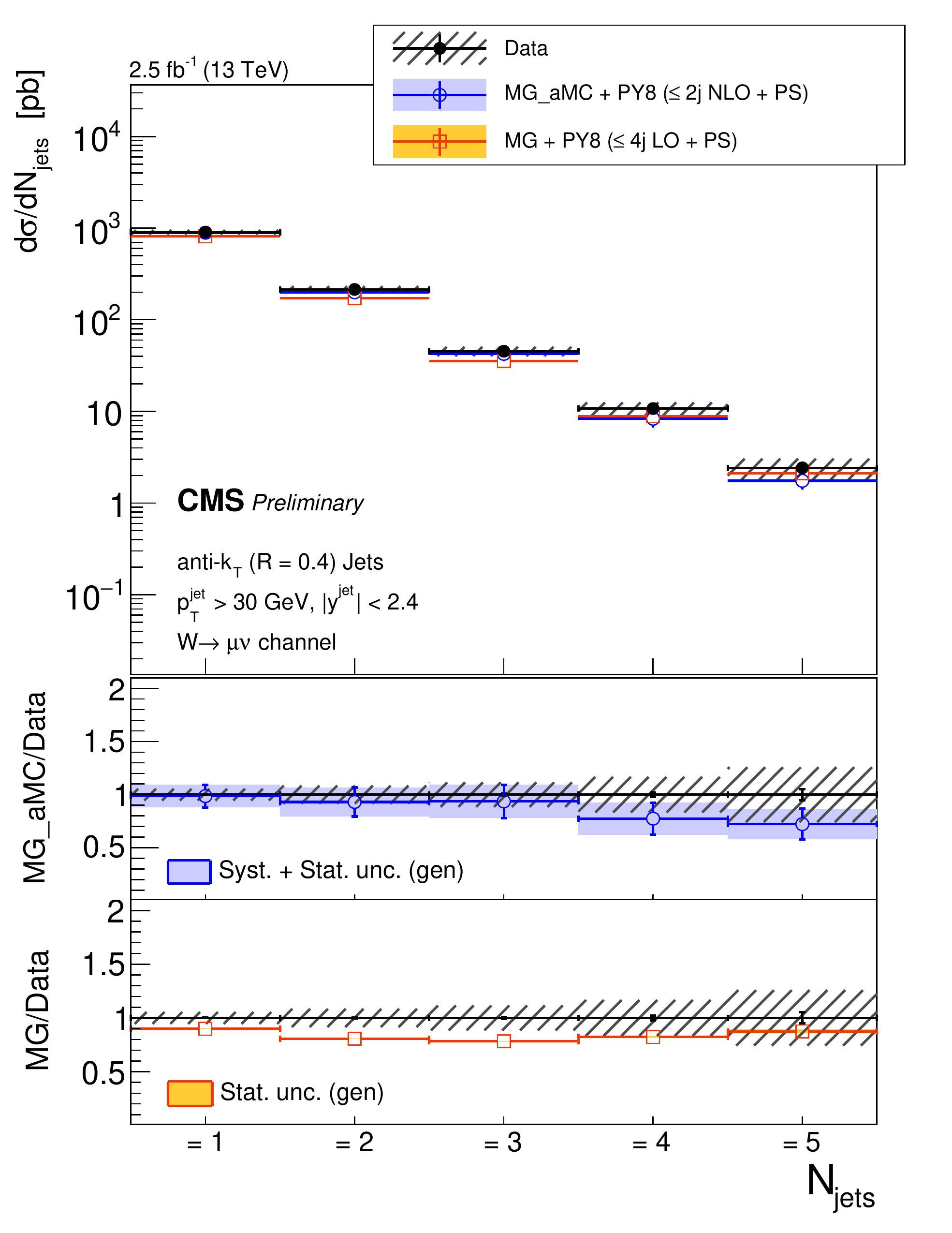}
\end{minipage}
\begin{minipage}[b]{0.32\textwidth}
\centering
\includegraphics[height=5.7cm]{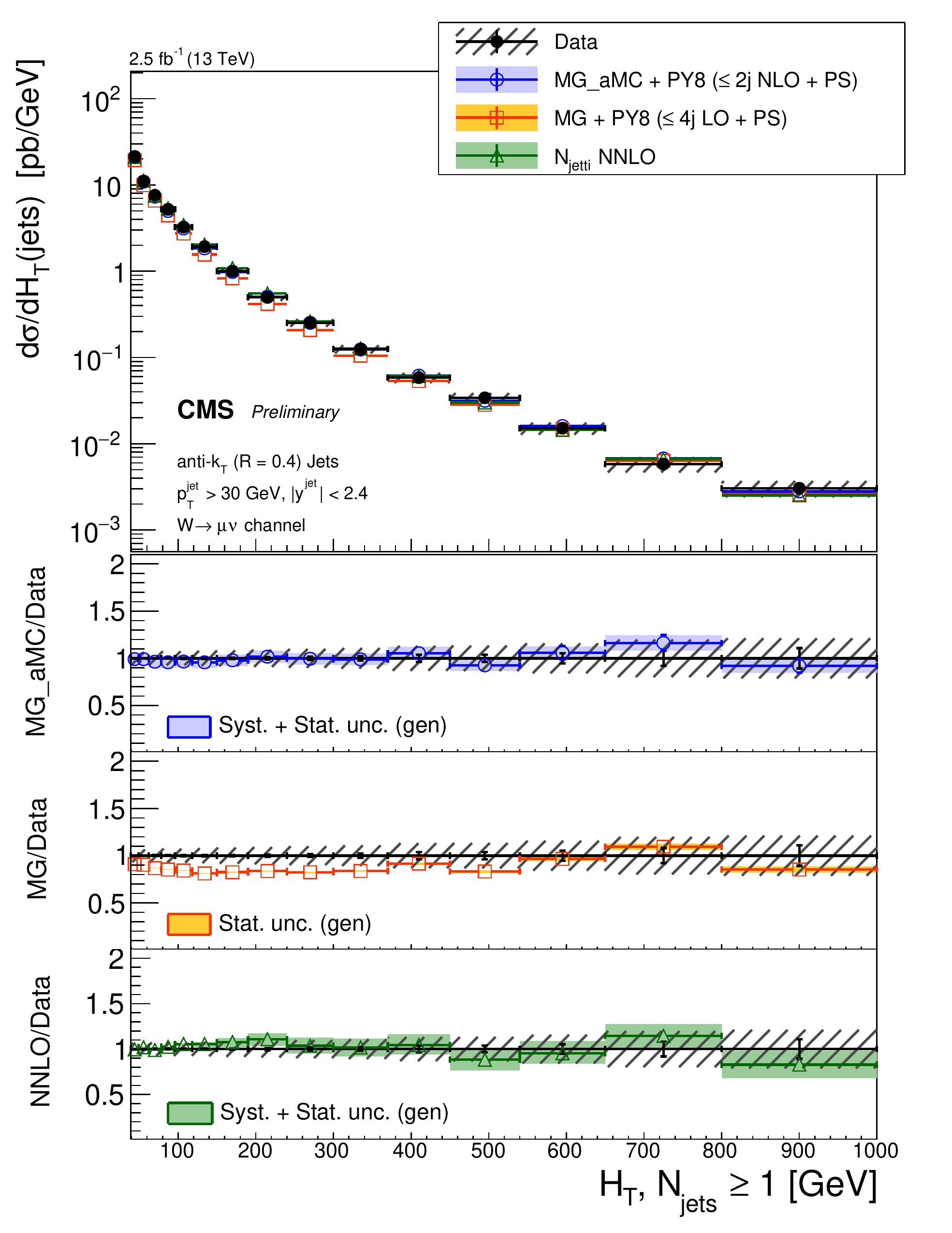}
\end{minipage}
\caption{Differential cross sections as a function of leading jet (left), exclusive jet multiplicities (middle) and $H_T$ for inclusive jet multiplicities 1 (right), compared to the predictions of~\MADGRAPH,~\MGvATNLO,~{\SHERPA 2},~\BLACKHAT{}+\SHERPA~\cite{wpjet:13tev}}.
\label{zwpjetfig2}
\end{figure}
\subsection{W+bb jets differential cross section at 8 TeV}
The production cross section of a~$\mathrm{W}(\rightarrow \ell\nu, where~\ell = e, \mu)$~boson with exactly two b jets with~$p_T > 25$~GeV and~$\eta < 2.4$~and no other jets 
with~$\eta < 4.7$, is measured using~$8~\mathrm{TeV}$ proton-proton collisions data corresponding to an integrated luminosity of $19.8~\mathrm{fb}^{-1}$~\cite{wpbb:8tev}. The 
results are also compared to theoretical prediction as shown in Fig.~\ref{zwpjetfig3}.
\begin{figure}[!ht]
\centering
\begin{minipage}[b]{0.45\textwidth}
\centering
\includegraphics[height=5.7cm]{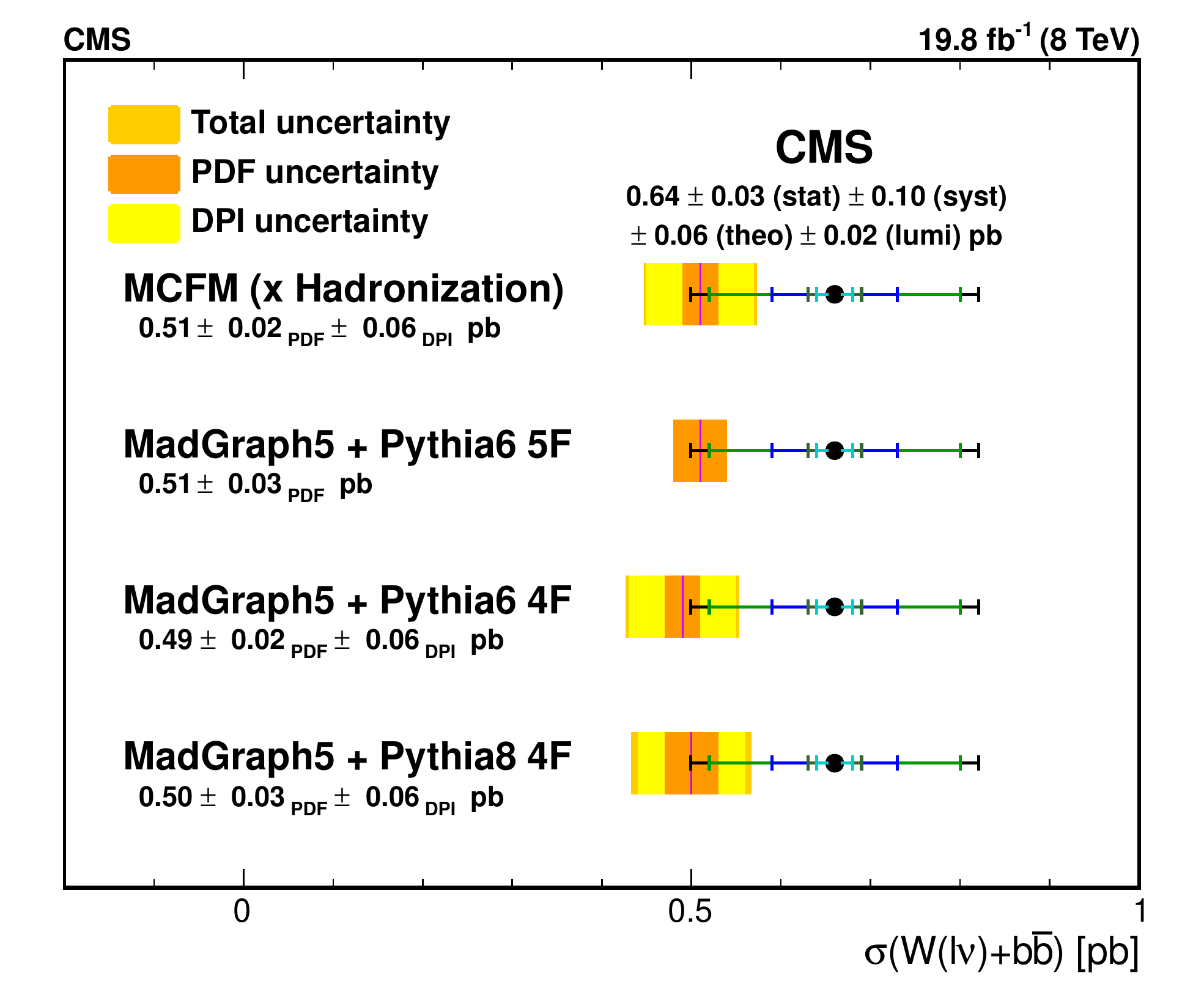}
\end{minipage}
\begin{minipage}[b]{0.45\textwidth}
\centering
\includegraphics[height=5.7cm]{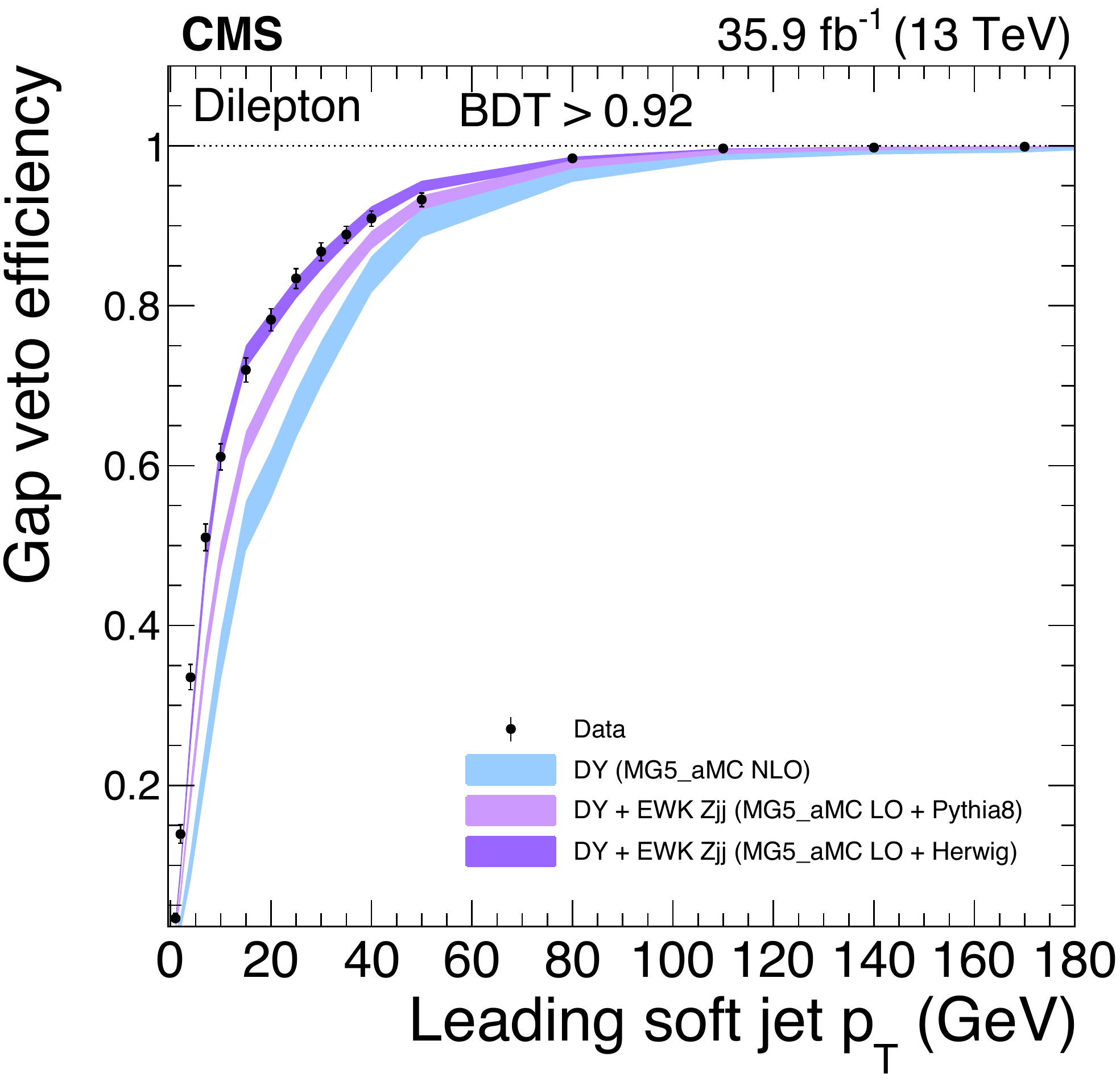}
\end{minipage}
\caption{Left:Comparison between the measured$\mathrm{W}(\rightarrow \ell\nu)+bb$ cross section and various QCD predictions. The orange band indicates the uncertainty in the given sample associated with PDF choice and the yellow band represents the uncertainty associated with DPI. The labels 4F and 5F refer to the four- and five-flavour PDF schemes~\cite{wpbb:8tev}. Right: Efficiency of a gap activity veto in dielectron and dimuon events with~$BDT > 0.92$, as a function of the leading soft jet~$p_{T}$~\cite{EWK:13tev}.}
\label{zwpjetfig3}
\end{figure}

\section{EWK production of Z+2 jets at 13 TeV}
The production of two electroweak jets (produced in hard interaction) in association with a Z($\rightarrow \ell\ell$, where~$\ell = e,~\mu$) boson is measured in proton-proton collision at~$\sqrt{13}$~TeV using data recorded by the CMS experiment corresponding to integrated luminosity of 35.9~$\ifb$~\cite{EWK:13tev}. The measurement is extracted in the kinematic region defined by~$M_{\ell\ell} > 50$~GeV,~$M_{{\rm jj}} > 120$~GeV and transverse momentum~$p_{T\rm{j}} > 25$~GeV. The cross section of the process is measured to be~$\sigma_{EW}(\ell\ell~{\rm jj}) = 552 \pm 19~({\rm stat.}) \pm 55~({\rm syst.})$~fb which is found to be in good agreement with SM prediction at leading order accuracy. Additionally, the associated jet activity of events in a signal-enriched region is also studied, and the measurements are found to be in agreement with QCD predictions, as shown in Fig.~\ref{zwpjetfig3}~(right).

\section{Conclusions}
CMS has made several SM V+Jets measurements using the LHC Run I and II datasets corresponding to integrated luminosities upto 19.8 (35.9)~$\ifb$~at~$\sqrt{s} = $~8 (13) TeV. All of the results are found to be consistent with the SM predictions. The SM will be tested with an unprecedented level of precision in new unexplored territories during the Run II, setting as well the ground for new physics.

\end{document}